\documentclass[twocolumn,showpacs,superscriptaddress,preprintnumbers,amsmath,amssymb,epsfig,floatfix,prx]{revtex4-1}
\setcitestyle{super}
\usepackage{graphicx}
\usepackage{gensymb}
\pdfoutput=1
\usepackage{amsmath}
\usepackage{color}
\usepackage{float}
\usepackage[english]{babel}
\usepackage{lipsum}

\usepackage{color,soul}
\bibpunct{[}{]}{,}{n}{}{}
\begin{document}
\title{High spin-polarization in the low Curie temperature complex itinerant ferromagnet EuTi$_{1-x}$Nb$_x$O$_3$}
\author{Suman Kamboj}
\affiliation{Department of Physical Sciences,
Indian Institute of Science Education and Research Mohali,
Sector 81, S. A. S. Nagar, Manauli, PO: 140306, India}
\author{Deepak K. Roy}
\affiliation{Department of Physics, Indian Institute of Science Education and Research, Pune 411008, India}

\author{Susmita Roy}
\affiliation{Condensed Matter Physics Division, Saha Institute of Nuclear Physics, Kolkata, PO: 700064, India}
\author{Rajeswari Roy Chowdhury}
\affiliation{Department of Physical Sciences,
Indian Institute of Science Education and Research Mohali,
Sector 81, S. A. S. Nagar, Manauli, PO: 140306, India}
\author{Prabhat Mandal}
\affiliation{Condensed Matter Physics Division, Saha Institute of Nuclear Physics, Kolkata, PO: 700064, India}
\author{Mukul Kabir}
\affiliation{Department of Physics, Indian Institute of Science Education and Research, Pune 411008, India}
\affiliation{Centre for Energy Science, Indian Institute of Science Education and Research, Pune 411008, India}
\author{Goutam Sheet}
\email {goutam@iisermohali.ac.in}
\affiliation{Department of Physical Sciences,
Indian Institute of Science Education and Research Mohali,
Sector 81, S. A. S. Nagar, Manauli, PO: 140306, India}

\date{\today}

\begin{abstract}
The physical systems with ferromagnetism and ``bad" metallicity hosting unusual transport properties are playgrounds of novel quantum phenomena. Recently EuTi$_{1-x}$Nb$_x$O$_3$ emerged as a ferromagnetic system where non-trivial temperature dependent transport properties are observed due to coexistence and competition of various magnetic and non-magnetic scattering processes. In the ferromagnetic state, the resistivity shows a $T^2$ temperature dependence possibly due to electron-
magnon scattering and above the Curie temperature $T_c$, the dependence changes to $T^{3/2}$ behaviour indicating a correlation between transport and magnetic properties.
In this paper, we show that the transport spin-polarization in EuTi$_{1-x}$Nb$_x$O$_3$, a low Curie temperature ferromagnet, is as high ($\sim 40\%$) as that in some of the metallic ferromagnets with high Curie temperatures. In addition, owing to the low Curie temperature of EuTi$_{1-x}$Nb$_x$O$_3$, the temperature ($T$) dependence of $P_t$ could be measured systematically up to $T_c$ which revealed a proportionate relationship with magnetization $M_s$ vs. $T$. This indicates that such proportionality is far more universally valid than the ferromagnets with ideal parabolic bands. Furthermore, our band structure calculations not only helped understand the origin of such high spin polarization in EuTi$_{1-x}$Nb$_x$O$_3$ but also provided a route to estimate the Hubbard $U$ parameter in complex metallic ferromagnets in general using experimental inputs.
\end{abstract}

\maketitle

Spintronic devices involve manipulation of the spin of electrons. Spin polarized electrons can, in principle, be sourced from itinerant ferromagnets with high degree of spin polarization\cite{Zutic2004,Prinz1995,Daughton1999}. In this regard, search, determination and understanding of spin polarization in novel materials is of utmost importance. Since device applications are desirable at ambient temperatures and most of the traditional ``transport" spin polarization measurement techniques rely on low-temperature experiments, it is most important to understand temperature evolution of the magnitude of spin polarization in materials. Such measurements are particularly interesting in materials that cannot be characterized by conventional metallicity. On the other hand, for a complete understanding of such temperature evolution up to the Curie temperature ($T_c$), it is important to have a ferromagnet where $T_c$ falls within the low-temperature limit where such measurements are possible. In this context, EuTi$_{1-x}$Nb$_x$O$_3$ is found to be the ideal candidate because it is a ``bad" metal and at the same time, a ferromagnet with low $T_c \sim$ 9.5 K for $x$ = 0.15\citep{PhysRevB.98.134428}.

EuTi$_{1-x}$Nb$_x$O$_3$ is an important member of the perovskite titanate family RTiO$_3$ (R = rare earth ion),where Ti usually has 3d$^1$ electronic configuration. This family has attracted researchers since last two decades owing to their several fascinating properties \cite{Zhao2015,Khaliullin2002,Hemberger2003,Ulrich2006}. There has been a considerable amount of theoretical and experimental work done for understanding the nature of the ground state in this family. Among the perovskites, EuTiO$_3$ with unique divalent rare-earth ion Eu and tetravalent Ti has generated substantial interest since the discovery of magnetoelectric coupling in this compound below the antiferromagnetic (AFM) transition temperature $T_N$ = 5.5 K \cite{Katsufuji1999,Katsufuji2001,Takahashi2009,Roy2016,Li2015,Li2014}. EuTiO$_3$ has a positive Curie- Weiss constant and exhibits strong coupling between magnetic ordering and phonon modes. Efforts have been made to drive this system from antiferromagnetic phase to other magnetic phases upon modifications due to partial substitution at the Ti site. Small amount of substitution at Ti site by Nb drives the system from antiferromagnetic to ferromagnetic \citep{1.4902137}. Substitution by Nb$^{4+}$ ion at the Ti$^{4+}$  results in the introduction of electrons in EuTiO$_3$ without disturbing the Eu$^{2+}$ moments chain. Previous studies  in EuTi$_{1−x}$Nb$_x$O$_3$ revealed notable increase in electrical conductivity  and large values of magnetocaloric response of about $23.8$ J kg$^{-1}$K$^{-1}$ even at low magnetic field 0-2 $T$ \cite{Roy2016}. Therefore, introduction of Nb in antiferromagnetic EuTiO$_3$ results in ferromagnetic behaviour in EuTi$_{1-x}$Nb$_x$O$_3$ ($x$ = 0.15) with T$_C\sim$ 9.5 K and saturation magnetization at 2 K is estimated to be $\sim$ 7$\mu_B$/Eu \cite{Roy2016}.

In this paper, we have employed spin-resolved Andreev reflection (PCAR) spectroscopy using conventional superconducting tip of Nb to measure the transport spin-polarization in EuTi$_{1-x}$Nb$_x$O$_3$ ($x$ = 0.15) \cite{Kamboj2018,Mukhopadhyay2007,Gupta,Chein,Leighton,Cohen,Mazin,Dhar,Grin}. Spin polarization ($P$) in a metallic ferromagnet is generally defined as $P$ = ({$N_\uparrow$(E$_F$)} - {$N_\downarrow$(E$_F$)/({$N_\uparrow$(E$_F$) + {$N_\downarrow$(E$_F$)), where $N_\uparrow$(E$_F$) and $N_\downarrow$(E$_F$) are the density of states (DOS) of the up and down spin channels respectively at the Fermi level. However, in a transport experiment like PCAR, the relevant quantity is not the absolute spin polarization but the so called ``transport spin polarization" which is defined as: $P_t$ = ({$\langle$$N_\uparrow$$v_{F\uparrow}$$\rangle$}$_{FS}$ - {$\langle$$N_\downarrow$$v_{F\downarrow}$$\rangle$}$_{FS}$)/({$\langle$$N_\uparrow$$v_{F\uparrow}$$\rangle$}$_{FS}$ + {$\langle$$N_\downarrow$$v_{F\downarrow}$$\rangle$}$_{FS}$) in the ballistic regime, where $N_\uparrow$ and $N_\downarrow$ are the density of states (DOS) of the up and down spin channels respectively at the Fermi level and $v_{F\uparrow}$ and $v_{F\downarrow}$ are the respective Fermi velocities \cite{Mazin1999}.

\begin{figure}[h!]
\includegraphics[width=8.5 cm]{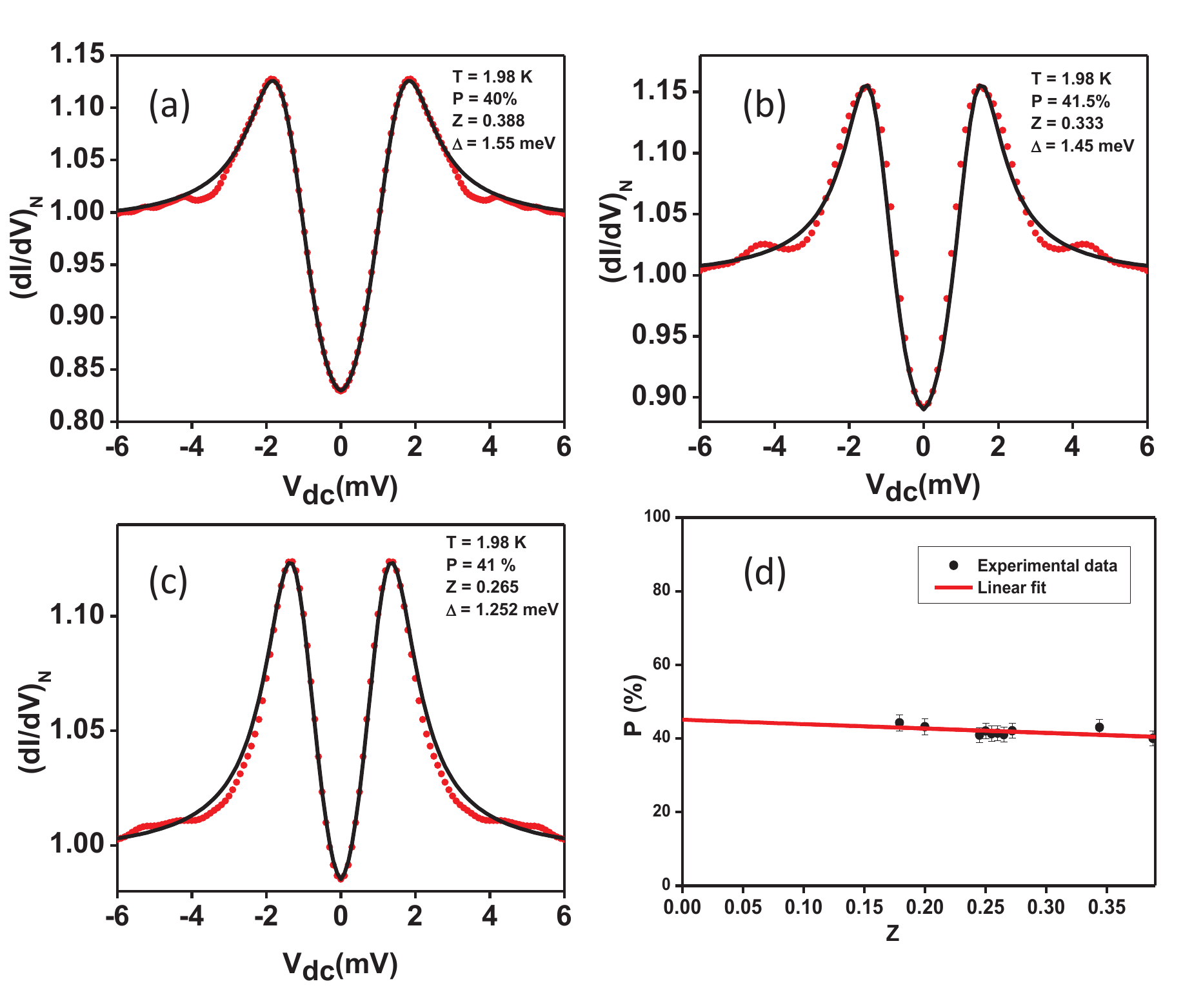}
\centering \caption[ ]
{\label{STM}   (a),(b),(c) Normalized dI/dV spectra for point-contacts on EuTi$_{1-x}$Nb$_x$O$_3$ with $x$=0.15 using a Nb tip. The black lines show BTK fits with spin-polarization included. (d) Spin-polarization (P$_t$) vs. barrier strength (Z) plot. The solid lines show extrapolation to Z = 0 where the spin-polarization approaches 45 \% }
\end{figure}

PCAR spectroscopy measurements were performed on single crystal EuTi$_{0.85}$Nb$_{0.15}$O$_3$ using Nb tips. The ballistic point-contacts between the sample and the tips were fabricated and controlled by moving the tip up and down by rotating a differential screw-based head assembly manually. The measurements involved obtaining the point contact spectra (i.e. differential conductance $dI/dV$ vs. $V$ curves) for different contacts with different values of $Z$, the interface transparency as in the Blonder-Tinkham-Klapwijk (BTK) theory. The spectra thus obtained were analyzed using BTK theory modified to incorporate the effect of spin polarized bands \cite{Strijkers2001,Blonder1982}
\begin{figure}[h!]
\includegraphics[width=8.5 cm]{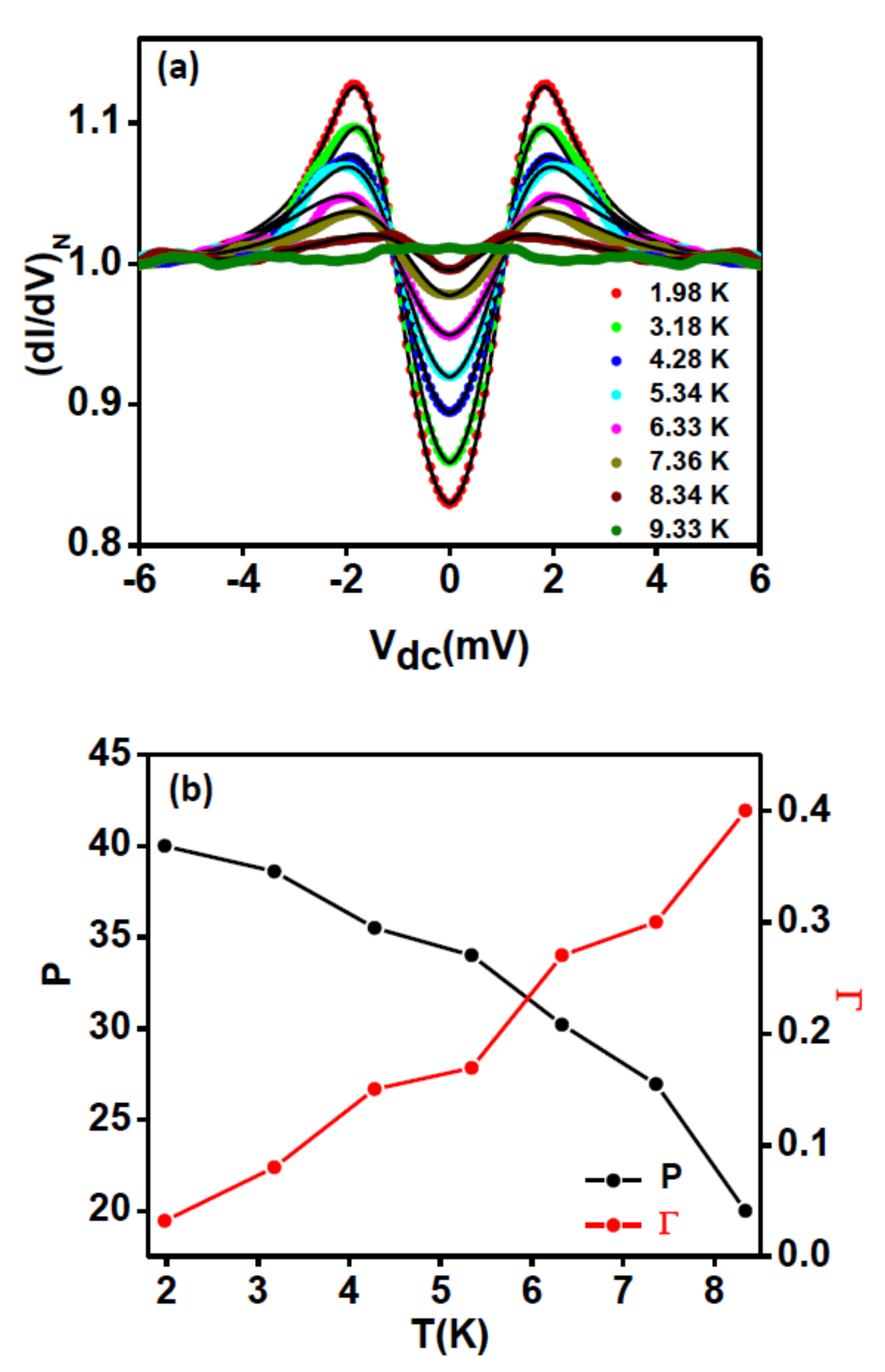}
\centering \caption[ ]
{\label{STM}  (a) Normalized dI/dV spectra with varying temperature over a temperature range of 1.98 K to 9.33 K for point-contacts on EuTi$_{0.85}$Nb$_{0.15}$O$_3$ using a Nb tip. The colored lines show experimental
data points and the black lines show BTK fits with spin-polarization included. (b) Temperature dependence of transport spin-polarization and the broadening parameter $\Gamma$. The solid lines are guide to the eye.}
\end{figure}

In Figure 1(a,b,c) we show three representative PCAR spectra (red points). The overall spectral features indicate that the point contacts belong to the ballistic or diffusive regime of transport where the two peaks symmetric about $V$ = 0 appear due to Andreev reflection across a superconducting interface with finite transparency ($Z$). The black lines show the fits to the experimentally obtained spectra using spin-polarized BTK theory. The superconducting gap $\Delta$ ranges between 1.2-1.5 meV for different contacts. The extracted values of spin-polarization $P_t$ and $\Gamma$ are also shown. For different values of $Z$ between 0.265 - 0.388 the transport spin polarization is measured to be around 41\%. The extracted values of $P_t$ is also seen to slightly depend (linearly) on $Z$ as shown in Figure 1(d). The solid lines in Figure 1(d) show linear extrapolation of the Z-dependence of $P_t$ to $Z$ = 0 which gives the expected intrinsic value of the spin-polarization (for $Z$ = 0). The intrinsic spin-polarization extracted in this case is found to be approximately 45\% which is comparable to elemental ferromagnetic metals like Fe ($P$ = 40\%), Co ($P$ = 42\%) and Nickel ($P$ = 39\%)\cite{Soulen1998}.

\begin{figure}[h!]
\includegraphics[width=8.5 cm]{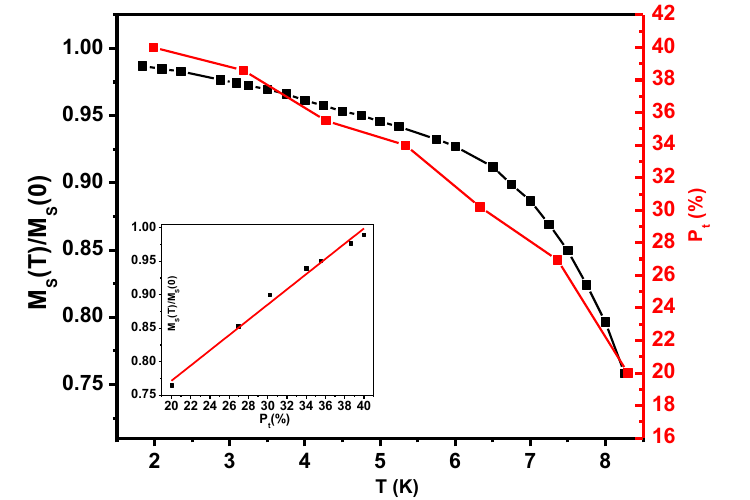}
\centering \caption[ ]
{\label{STM}  Temperature dependence of transport spin-polarization ($P$) and spontaneous magnetization M$_S$(T) normalized by that at T=0 K,in the temperature range 2 - 9 K. Inset shows variation of normalized magnetization with transport spin-polarization. The solid lines are guide to the eye.}
\end{figure}

\begin{figure*}[!t]
\begin{center}
\rotatebox{0}{\includegraphics[width=0.98\textwidth]{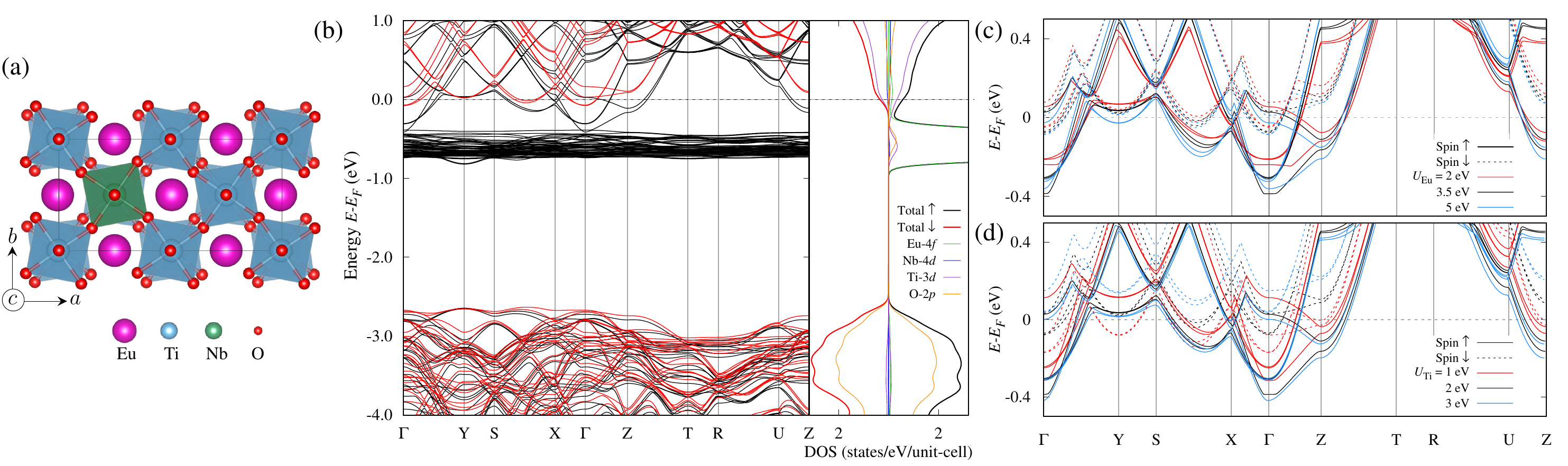}}
\caption{(a) Crystal structure of EuTi$_{1-x}$Nb$_x$O$_3$ indicates the oxygen octahedra tilting. (b)  Band structure and the corresponding density of states for $x$=0.125. Majority (minority) spin channel is drawn with black (red) color. The localized Eu-4$f$ states lie about 0.5 eV below the Fermi level. The Fermi level is composed of Ti-3$d$, Eu-4$f$ and Nb-4$d$ electrons. (c)-(d) Energy evolution of the low-energy bands with Hubbard $U$ parameter. While $U_{\rm Eu}$ is varied with fixed $U_{\rm Ti}$=2 eV in (c), the evolution of bands with different $U_{\rm Ti}$ ($U_{\rm Eu}$ = 3.5 eV) is shown in (d).}
\label{fig:figure-t}
\end{center}
\end{figure*}
Figure 2(a) shows temperature dependence of normalized dI/dV spectra for point-contacts on EuTi$_{0.85}$Nb$_{0.15}$O$_3$. The dip at $V$ = 0 in the spectra is maximum at the lowest observed temperature of 1.98 K. With increase in temperature, the differential conductance spectra gradually undergo thermal smearing. At temperatures greater than 8.34 K the most dominant feature of the spectra, namely, the two peaks in the conductance spectra associated with the superconducting energy gap gets flattened.

The BTK fits of the temperature dependence of normalized $dI/dV$ spectra (solid line in Figure 2(a)) give an estimate of temperature dependence of the measured transport spin-polarization $P_t$ and also the broadening parameter $\Gamma$ as shown in Figure 2(b). Transport spin-polarization decreases with increase in temperature whereas the parameter $\Gamma$ is observed to increase with temperature. $\Gamma$ gradually increases and reaches a maximum value of 0.4meV at 8 K close to the T$_C$ of EuTi$_{0.85}$Nb$_{0.15}$O$_3$. The parameter $\Gamma$ is associated with lifetime ($\tau$) of superconducting quasiparticle in the way $\Gamma$ = $\hbar$/$\tau$. The increase in $\Gamma$ with temperature signifies a corresponding decrease in quasiparticle lifetime at the same temperature which might be due to enhanced spin fluctuations at higher temperatures. Figure 3 shows temperature dependence of transport spin-polarization and spontaneous magnetization in the same panel for a direct comparison between the two. $P_t$ is observed to decrease with temperature following the same behaviour as $M_s$. This is clearly seen in the inset of Figure 3 where $M_s$ and $P_t$ are seen to linearly dependent on each other. This is remarkable because such proportional dependence of $P_t$ and $M_s$ with temperature is expected only for materials with strictly parabolic bands and EuTi$_{0.85}$Nb$_{0.15}$O$_3$ is known to be a system with large correlation effects\cite{Mukhopadhyay2007}.

To gain insight into the experimental results, we have performed first-principles  calculations within the density functional theory as implemented in the VASP.~\citep{PhysRev.136.B864,PhysRev.140.A1133,PhysRevB.47.558,PhysRevB.54.11169} The spin-polarized wavefunctions are described within the projector augmented wave formalism with 550 eV cutoff for the kinetic energy.~\citep{PhysRevB.50.17953} The exchange-correlation energy is described with the Perdew-Burke-Ernzerhof functional.~\citep{PhysRevLett.77.3865} The strong electron correlation in Eu-$4f$, Ti-$3d$, and Nb-$4d$ electrons is accounted by an on-site Hubbard-type Coulomb interaction $U$ within the rotationally invariant Dudarev's approach.~\citep{PhysRevB.57.1505}  While a $\sqrt{2}\times\sqrt{2}\times 2$ tetragonal supercell is used to determine the correct magnetic ground state of pure EuTiO$_3$, a $2\sqrt{2}\times\sqrt{2}\times 2$ orthorhombic supercell is used  to determine the electronic structure and spin transport polarization in EuTi$_{1-x}$Nb$_x$O$_3$.  The Brillouin zone is sampled with 4$\times$8$\times$6 k-point mesh according to the Monkhorst-Pack scheme,~\citep{PhysRevB.13.5188} while the atomic positions, volume and the shape of the supercell are optimized until all the force components are less than 0.01 eV/\AA~ threshold. The calculated lattice parameter of 3.97 \AA~ for EuTiO$_3$ compares well with the experimental value of 3.91 \AA.~\cite{Katsufuji2001}

In agreement with the previous experimental and theoretical results, the present DFT+$U$ ($U_{\rm Eu}$ = 3.5 and  $U_{\rm Ti}$ = 2 eV) calculations predict an antiferromagnetic (AFM) insulating ground state for EuTiO$_3$.~\citep{1.1708549,Lee2010}  The AFM superexchange between the half-filled 4$f$ states of Eu$^{2+}$ ($S$=7/2) via the Ti-$3d$ states wins over the ferromagnetic (FM) indirect exchange via Eu-$5d$ states.~\citep{PhysRevB.83.214421} The competing FM state lies only 1.2 meV/f.u. higher in energy. Thus, within the DFT+$U$ approach, the tuneable Hubbard $U$ parameter dictates the long-range magnetic ordering.~\citep{Ranjan_2007,PhysRevB.88.094103} In the absence of an on-site Coulomb interaction for the Ti-$3d$ electrons, the $U_{\rm Eu}$ above 5 eV favours the competing FM state. In contrast, while $U_{\rm Ti}$ = 2 eV is used, a lower $U_{\rm Eu} < $ 4 eV describes the AFM ground state correctly. Such decrease in $U_{\rm Eu}$ with an onset of $U$ for the Ti-$3d$ electrons is suggestive, since, if the Ti electrons are assumed to be more localized then the Eu electrons need to be more delocalized in order to maintain the superexchange interaction described above. The long-range AFM order can be tuned to FM through strain, pressure, and chemical doping.~\citep{Lee2010,1.4902137} It is known that upon doping EuTiO$_3$ with more than 5\% Nb in place of Ti in EuTi$_{1-x}$Nb$_x$O$_3$, the magnetic structure changes from AFM to FM.~\citep{1.4902137} The Nb atoms in the lattice introduce one itinerant electron per Ti-atom replaced, and as a result, EuTi$_{1-x}$Nb$_x$O$_3$ becomes metallic. The itinerant Nb-4$d$ electrons mediate the Ruderman-Kittel-Kasuya-Yoshida exchange interaction between the Eu$^{2+}$ ions, and the FM ground state emerges.~\citep{PhysRevB.98.134428}

The EuTi$_{1-x}$Nb$_x$O$_3$ with $x$ = 0.15  ($U_{\rm Eu}$ = 3.5 and  $U_{\rm Ti/Nb}$ = 2 eV) shown in Figure~\ref{fig:figure-t}(a) becomes FM metal [Figure~\ref{fig:figure-t}(b)], and we estimate the transport spin polarization in the ballistic limit. 
The magnetic moment of 6.85 $\mu_B$ at the Eu-site is in agreement with an earlier experiment,~\citep{1.1708549}  which arises from the localized Eu-4$f$ orbitals [Figure~\ref{fig:figure-t}(b)], which are around 0.5 eV below the Fermi level. A deeper investigation into the density of states (DOS) indicates that the states at the Fermi level $E_F$ is composed of Ti-3$d$ states with small contributions from the Eu-4$f$ and Nb-4$d$ states [Figure~\ref{fig:figure-t}(b)]. The spin-polarized DOS at the $E_F$ is calculated to be $N_{\uparrow}(E_F)$ = 0.574 states/eV/unit-cell which is 68\% higher than the $N_{\downarrow}(E_F)$. The average Fermi velocity $\langle v_{F\uparrow} \rangle$ = 3.47$\times$10$^5$ m/s is also found to be higher for the spin-up channel when compared to the spin-down channel, $\langle v_{F\downarrow} \rangle$ = 1.99$\times$10$^5$ m/s.  These result in a very high transport spin polarization of 49\% in the ballistic regime, which is in excellent agreement with the present experimental results with $x$ = 0.15.

It is important to discuss that the electronic structure and the concurrent $P_t$ crucially depends on the on-site Coulomb interaction $U_{\rm Eu/Ti/Nb}$, which is treated as a parameter within the DFT+$U$ formalism.~\citep{PhysRevB.44.943} We find that with increasing $U_{\rm Eu}$ [Figure~\ref{fig:figure-t}(c)] both the channels are pushed down in energy and the corresponding $P_t$ monotonically decreases to reach 25\%  for $U_{\rm Eu}$ = 5 eV, while the $U_{\rm Ti}$ is kept fixed at 2 eV. In contrast, with increasing $U_{\rm Ti}$ [Figure~\ref{fig:figure-t}(d)], the minority channel is pushed higher in energy and the corresponding $P_t$ increases monotonically. Ultimately a half-metallic solution emerges for $U_{\rm Ti}$ = 3 eV with 100\% spin polarization. Thus, in the absence of a first-principles estimation of Hubbard $U$, one can use the experimental knowledge of transport spin polarization to better estimate the theoretical $U$ parameter in ferromagnetic metals.

In conclusion, we report point contact Andreev reflection spectroscopy on polycrystalline of EuTi$_{0.85}$Nb$_{0.15}$O$_3$. EuTi$_{0.85}$Nb$_{0.15}$O$_3$ is ferromagnetic with Curie temperature T$_C$ = 9.5 K. Transport spin polarization ($P_t$) estimated from differential conductance spectra is about 45\% which is close to elemental ferromagnets. Temperature dependence of transport spin polarization closely follows the temperature dependence of spontaneous magnetization ($M_S$). Furthermore, temperature dependence of the broadening parameter ($\Gamma$) estimated from PCAR spectra exhibits increase with temperature. This implies with increase in temperature superconducting quasiparticle lifetime decreases. This decrease can be attributed to the effect of spin fluctuations close to the critical temperature of the ferromagnetic system. Thus, our study can provide an alternative way to probe spin fluctuations in ferromagnetic materials. Based on a comparative analysis between the experimental and theoretical investigations we have also discussed a possible route to directly estimate the Hubbard $U$ parameter in complex metallic ferromagnets in general.

GS acknowledges financial support from the research grants of (a) Swarnajayanti fellowship awarded by the Department of Science and Technology (DST), Govt. of India under the grant number \textbf{DST/SJF/PSA-01/2015-16} and (b) from SERB under the grant number \textbf{EMR/2015/001650}. M. K. acknowledges funding from the Science and Engineering Research Board through Nano Mission project \textbf{SR/NM/TP-13/2016}.


\begin{thebibliography}{17}%
\makeatletter
\providecommand \@ifxundefined [1]{%
 \@ifx{#1\undefined}
}%
\providecommand \@ifnum [1]{%
 \ifnum #1\expandafter \@firstoftwo
 \else \expandafter \@secondoftwo
 \fi
}%
\providecommand \@ifx [1]{%
 \ifx #1\expandafter \@firstoftwo
 \else \expandafter \@secondoftwo
 \fi
}%
\providecommand \natexlab [1]{#1}%
\providecommand \enquote  [1]{``#1''}%
\providecommand \bibnamefont  [1]{#1}%
\providecommand \bibfnamefont [1]{#1}%
\providecommand \citenamefont [1]{#1}%
\providecommand \href@noop [0]{\@secondoftwo}%
\providecommand \href [0]{\begingroup \@sanitize@url \@href}%
\providecommand \@href[1]{\@@startlink{#1}\@@href}%
\providecommand \@@href[1]{\endgroup#1\@@endlink}%
\providecommand \@sanitize@url [0]{\catcode `\\12\catcode `\$12\catcode
  `\&12\catcode `\#12\catcode `\^12\catcode `\_12\catcode `\%12\relax}%
\providecommand \@@startlink[1]{}%
\providecommand \@@endlink[0]{}%
\providecommand \url  [0]{\begingroup\@sanitize@url \@url }%
\providecommand \@url [1]{\endgroup\@href {#1}{\urlprefix }}%
\providecommand \urlprefix  [0]{URL }%
\providecommand \Eprint [0]{\href }%
\providecommand \doibase [0]{http://dx.doi.org/}%
\providecommand \selectlanguage [0]{\@gobble}%
\providecommand \bibinfo  [0]{\@secondoftwo}%
\providecommand \bibfield  [0]{\@secondoftwo}%
\providecommand \translation [1]{[#1]}%
\providecommand \BibitemOpen [0]{}%
\providecommand \bibitemStop [0]{}%
\providecommand \bibitemNoStop [0]{.\EOS\space}%
\providecommand \EOS [0]{\spacefactor3000\relax}%
\providecommand \BibitemShut  [1]{\csname bibitem#1\endcsname}%
\let\auto@bib@innerbib\@empty


\bibitem{Zutic2004} I. Zutic, J. Fabian, and S. Das Sarma, Rev. Mod. Phys. \textbf{76}, 323 (2004).
\bibitem{Prinz1995} G. A. Prinz, Phys. Today \textbf{48}, 58 (1995).
\bibitem{Daughton1999} J. M. Daughton, A. V. Pohm, R. T. Fayfield, and C. H. Smith, J. Phys. D \textbf{32}, R169 (1999).



\bibitem{Zhao2015} Z. Y. Zhao, O. Khosravani, M. Lee, L. Balicas, X. F. Sun, J. G. Cheng, J. Brooks, H. D. Zhou, and E. S. Choi, Phys. Rev.B \textbf{91}, 161106(R) (2015).

\bibitem{Khaliullin2002} G. Khaullin and S. Okamoto, Phys. rev. lett. \textbf{89},167201 (2002).

\bibitem{Hemberger2003} J.Hemberger, H.-A. Krug von Nidda, V. Fritsch, J. Deisenhofer, S. Lobina, T. Rudolf, P. Lunkenheimer, F. Lichtenberg, A.Loidl, D. Bruns, and B. Büchner, Phys. Rev. Lett. \textbf{91}, 066403 (2003).

\bibitem{Ulrich2006}  C. Ulrich, A. Gössling, M. Grüninger, M. Guennou, H. Roth,M. Cwik, T. Lorenz, G. Khaliullin, and B. Keimer,Phys. Rev.Lett. \textbf{97},157401 (2006)


\bibitem{Katsufuji1999} T. Katsufuji and Y. Tokura,Phys.Rev.B \textbf{60},R15021 (1999).

\bibitem{Katsufuji2001} T. Katsufuji and H. Takagi,Phys.Rev.B \textbf{64},054415 (2001).

\bibitem{Takahashi2009} K. S. Takahashi, M. Onoda, M. Kawasaki, N. Nagaosa, and Y.Tokura, Phys.Rev.Lett. \textbf{103},057204 (2009).

\bibitem{Roy2016} S. Roy, N. Khan, and P. Mandal,APL Mater.\textbf{4},026102 (2016).


\bibitem{Li2015} L. Li, J. R. Morris, M. R. Koehler, Z. Dun, H. Zhou, J. Yan, D.Mandrus, and V. Keppens,Phys. Rev. B \textbf{92},024109 (2015).

\bibitem{Li2014} L. Li, H. Zhou, J. Yan, D. Mandrus, and V. Keppens,APLMater. \textbf{2},110701 (2014).

\bibitem{Kamboj2018} S. Kamboj, S. Das, A. Sirohi, R. R. Chowdhury, S. Gayen, V. K. Maurya, S. Patnaik and G. Sheet, J. Phys.:  Condens. Matter \textbf{30}, 355001 (2018).


\bibitem{Mukhopadhyay2007} S. Mukhopadhyay, P. Raychaudhuri, D. A. Joshi and C.V. Tomy, Phys. Rev. B \textbf{75}, 014504 (2007).


\bibitem{Gupta} Y. Ji, G. J. Strijkers, F. Y. Yang, C. L. Chien, J. M. Byers, A.Anguelouch, G. Xiao, and A. Gupta, Phys. Rev. Lett. \textbf{86}, 5585 (2001).

\bibitem{Chein} J. M. Valentine and C. L. Chien, J. Appl. Phys. \textbf{99}, 08P902 (2006).

\bibitem{Leighton} L. Wang, K. Umemoto, R. M. Wentzcovitch, T. Y. Chen, C. L.Chien, J. G. Checkelsky, J. C. Eckert, E. D. Dahlberg, and C.Leighton, Phys. Rev. Lett. \textbf{94}, 056602 (2005).

\bibitem{Cohen} S. K. Clowes, Y. Miyoshi, Y. Bugoslavsky, W. R. Branford, C.Grigorescu, S. A. Manea, O. Monnereau, and L. F. Cohen, Phys.Rev. B \textbf{69}, 214425 (2004).

\bibitem{Mazin} R. Panguluri, G. Tsoi, B. Nadgorny, S. H. Chun, N. Samarth, andI. I. Mazin, Phys. Rev. B \textbf{68}, 201307(R) (2003).

\bibitem{Dhar} S. Singh, G. Sheet, P. Raychaudhuri, and S. K. Dhar, Appl. Phys.Lett. \textbf{88}, 022506 (2006).

\bibitem{Grin} G. Sheet, H. Rosner, S. Wirth, A.Leithe-Jasper, W. Schnelle, U. Burkhardt, J. A. Mydosh, P. Ray-chaudhuri, and Y. Grin, Phys. Rev. B \textbf{72}, 180407 (R) (2005).

\bibitem{Mazin1999} I. I. Mazin,Phys. Rev. Lett. \textbf{83}, 1427 (1999).

\bibitem{Strijkers2001} G. J. Strijkers, Y. Ji, F. Y. Yang, C. L. Chien, and J. M.Byers, Phys. Rev. B\textbf{63}, 104510 (2001).

\bibitem{Blonder1982} G. E. Blonder, M. Tinkham, and T. M. Klapwijk, Phys.Rev. B \textbf{25}, 4515 (1982)


\bibitem{Soulen1998} R. J. Soulen Jr., J. M. Byers, M. S. Osofsky, B. Nad-gorny, T. Ambrose, S. F. Cheng, P. R. Broussard, C. T.Tanaka, J. Nowak, J. S. Moodera, A. Barry, J. M. D.Coey, Science \textbf{282}, 85 (1998).





\bibitem [{\citenamefont {Hohenberg}\ and\ \citenamefont
  {Kohn}(1964)}]{PhysRev.136.B864}%
  \BibitemOpen
  \bibfield  {author} {\bibinfo {author} {\bibfnamefont {P.}~\bibnamefont
  {Hohenberg}}\ and\ \bibinfo {author} {\bibfnamefont {W.}~\bibnamefont
  {Kohn}},\ }\href {\doibase 10.1103/PhysRev.136.B864} {\bibfield  {journal}
  {\bibinfo  {journal} {Phys. Rev.}\ }\textbf {\bibinfo {volume} {136}},\
  \bibinfo {pages} {B864} (\bibinfo {year} {1964})}\BibitemShut {NoStop}%
\bibitem [{\citenamefont {Kohn}\ and\ \citenamefont
  {Sham}(1965)}]{PhysRev.140.A1133}%
  \BibitemOpen
  \bibfield  {author} {\bibinfo {author} {\bibfnamefont {W.}~\bibnamefont
  {Kohn}}\ and\ \bibinfo {author} {\bibfnamefont {L.~J.}\ \bibnamefont
  {Sham}},\ }\href {\doibase 10.1103/PhysRev.140.A1133} {\bibfield  {journal}
  {\bibinfo  {journal} {Phys. Rev.}\ }\textbf {\bibinfo {volume} {140}},\
  \bibinfo {pages} {A1133} (\bibinfo {year} {1965})}\BibitemShut {NoStop}%
\bibitem [{\citenamefont {Kresse}\ and\ \citenamefont
  {Hafner}(1993)}]{PhysRevB.47.558}%
  \BibitemOpen
  \bibfield  {author} {\bibinfo {author} {\bibfnamefont {G.}~\bibnamefont
  {Kresse}}\ and\ \bibinfo {author} {\bibfnamefont {J.}~\bibnamefont
  {Hafner}},\ }\href {\doibase 10.1103/PhysRevB.47.558} {\bibfield  {journal}
  {\bibinfo  {journal} {Phys. Rev. B}\ }\textbf {\bibinfo {volume} {47}},\
  \bibinfo {pages} {558} (\bibinfo {year} {1993})}\BibitemShut {NoStop}%
\bibitem [{\citenamefont {Kresse}\ and\ \citenamefont
  {Furthm\"uller}(1996)}]{PhysRevB.54.11169}%
  \BibitemOpen
  \bibfield  {author} {\bibinfo {author} {\bibfnamefont {G.}~\bibnamefont
  {Kresse}}\ and\ \bibinfo {author} {\bibfnamefont {J.}~\bibnamefont
  {Furthm\"uller}},\ }\href {\doibase 10.1103/PhysRevB.54.11169} {\bibfield
  {journal} {\bibinfo  {journal} {Phys. Rev. B}\ }\textbf {\bibinfo {volume}
  {54}},\ \bibinfo {pages} {11169} (\bibinfo {year} {1996})}\BibitemShut
  {NoStop}%
\bibitem [{\citenamefont {Bl\"ochl}(1994)}]{PhysRevB.50.17953}%
  \BibitemOpen
  \bibfield  {author} {\bibinfo {author} {\bibfnamefont {P.~E.}\ \bibnamefont
  {Bl\"ochl}},\ }\href {\doibase 10.1103/PhysRevB.50.17953} {\bibfield
  {journal} {\bibinfo  {journal} {Phys. Rev. B}\ }\textbf {\bibinfo {volume}
  {50}},\ \bibinfo {pages} {17953} (\bibinfo {year} {1994})}\BibitemShut
  {NoStop}%
\bibitem [{\citenamefont {Perdew}\ \emph {et~al.}(1996)\citenamefont {Perdew},
  \citenamefont {Burke},\ and\ \citenamefont
  {Ernzerhof}}]{PhysRevLett.77.3865}%
  \BibitemOpen
  \bibfield  {author} {\bibinfo {author} {\bibfnamefont {J.~P.}\ \bibnamefont
  {Perdew}}, \bibinfo {author} {\bibfnamefont {K.}~\bibnamefont {Burke}}, \
  and\ \bibinfo {author} {\bibfnamefont {M.}~\bibnamefont {Ernzerhof}},\ }\href
  {\doibase 10.1103/PhysRevLett.77.3865} {\bibfield  {journal} {\bibinfo
  {journal} {Phys. Rev. Lett.}\ }\textbf {\bibinfo {volume} {77}},\ \bibinfo
  {pages} {3865} (\bibinfo {year} {1996})}\BibitemShut {NoStop}%
\bibitem [{\citenamefont {Dudarev}\ \emph {et~al.}(1998)\citenamefont
  {Dudarev}, \citenamefont {Botton}, \citenamefont {Savrasov}, \citenamefont
  {Humphreys},\ and\ \citenamefont {Sutton}}]{PhysRevB.57.1505}%
  \BibitemOpen
  \bibfield  {author} {\bibinfo {author} {\bibfnamefont {S.~L.}\ \bibnamefont
  {Dudarev}}, \bibinfo {author} {\bibfnamefont {G.~A.}\ \bibnamefont {Botton}},
  \bibinfo {author} {\bibfnamefont {S.~Y.}\ \bibnamefont {Savrasov}}, \bibinfo
  {author} {\bibfnamefont {C.~J.}\ \bibnamefont {Humphreys}}, \ and\ \bibinfo
  {author} {\bibfnamefont {A.~P.}\ \bibnamefont {Sutton}},\ }\href {\doibase
  10.1103/PhysRevB.57.1505} {\bibfield  {journal} {\bibinfo  {journal} {Phys.
  Rev. B}\ }\textbf {\bibinfo {volume} {57}},\ \bibinfo {pages} {1505}
  (\bibinfo {year} {1998})}\BibitemShut {NoStop}%
\bibitem [{\citenamefont {Monkhorst}\ and\ \citenamefont
  {Pack}(1976)}]{PhysRevB.13.5188}%
  \BibitemOpen
  \bibfield  {author} {\bibinfo {author} {\bibfnamefont {H.~J.}\ \bibnamefont
  {Monkhorst}}\ and\ \bibinfo {author} {\bibfnamefont {J.~D.}\ \bibnamefont
  {Pack}},\ }\href {https://link.aps.org/doi/10.1103/PhysRevB.13.5188}
  {\bibfield  {journal} {\bibinfo  {journal} {Phys. Rev. B}\ }\textbf {\bibinfo
  {volume} {13}},\ \bibinfo {pages} {5188} (\bibinfo {year}
  {1976})}\BibitemShut {NoStop}%
\bibitem [{\citenamefont {McGuire}\ \emph {et~al.}(1966)\citenamefont
  {McGuire}, \citenamefont {Shafer}, \citenamefont {Joenk}, \citenamefont
  {Alperin},\ and\ \citenamefont {Pickart}}]{1.1708549}%
  \BibitemOpen
  \bibfield  {author} {\bibinfo {author} {\bibfnamefont {T.~R.}\ \bibnamefont
  {McGuire}}, \bibinfo {author} {\bibfnamefont {M.~W.}\ \bibnamefont {Shafer}},
  \bibinfo {author} {\bibfnamefont {R.~J.}\ \bibnamefont {Joenk}}, \bibinfo
  {author} {\bibfnamefont {H.~A.}\ \bibnamefont {Alperin}}, \ and\ \bibinfo
  {author} {\bibfnamefont {S.~J.}\ \bibnamefont {Pickart}},\ }\href {\doibase
  10.1063/1.1708549} {\bibfield  {journal} {\bibinfo  {journal} {J. Appl.
  Phys.}\ }\textbf {\bibinfo {volume} {37}},\ \bibinfo {pages} {981} (\bibinfo
  {year} {1966})}\BibitemShut {NoStop}%
\bibitem [{\citenamefont {Lee}\ \emph {et~al.}(2010)\citenamefont {Lee},
  \citenamefont {Fang}, \citenamefont {Vlahos}, \citenamefont {Ke},
  \citenamefont {Jung}, \citenamefont {Kourkoutis}, \citenamefont {Kim},
  \citenamefont {Ryan}, \citenamefont {Heeg}, \citenamefont {Roeckerath},
  \citenamefont {Goian}, \citenamefont {Bernhagen}, \citenamefont {Uecker},
  \citenamefont {Hammel}, \citenamefont {Rabe}, \citenamefont {Kamba},
  \citenamefont {Schubert}, \citenamefont {Freeland}, \citenamefont {Muller},
  \citenamefont {Fennie}, \citenamefont {Schiffer}, \citenamefont {Gopalan},
  \citenamefont {Johnston-Halperin},\ and\ \citenamefont {Schlom}}]{Lee2010}%
  \BibitemOpen
  \bibfield  {author} {\bibinfo {author} {\bibfnamefont {J.~H.}\ \bibnamefont
  {Lee}}, \bibinfo {author} {\bibfnamefont {L.}~\bibnamefont {Fang}}, \bibinfo
  {author} {\bibfnamefont {E.}~\bibnamefont {Vlahos}}, \bibinfo {author}
  {\bibfnamefont {X.}~\bibnamefont {Ke}}, \bibinfo {author} {\bibfnamefont
  {Y.~W.}\ \bibnamefont {Jung}}, \bibinfo {author} {\bibfnamefont {L.~F.}\
  \bibnamefont {Kourkoutis}}, \bibinfo {author} {\bibfnamefont {J.-W.}\
  \bibnamefont {Kim}}, \bibinfo {author} {\bibfnamefont {P.~J.}\ \bibnamefont
  {Ryan}}, \bibinfo {author} {\bibfnamefont {T.}~\bibnamefont {Heeg}}, \bibinfo
  {author} {\bibfnamefont {M.}~\bibnamefont {Roeckerath}}, \bibinfo {author}
  {\bibfnamefont {V.}~\bibnamefont {Goian}}, \bibinfo {author} {\bibfnamefont
  {M.}~\bibnamefont {Bernhagen}}, \bibinfo {author} {\bibfnamefont
  {R.}~\bibnamefont {Uecker}}, \bibinfo {author} {\bibfnamefont {P.~C.}\
  \bibnamefont {Hammel}}, \bibinfo {author} {\bibfnamefont {K.~M.}\
  \bibnamefont {Rabe}}, \bibinfo {author} {\bibfnamefont {S.}~\bibnamefont
  {Kamba}}, \bibinfo {author} {\bibfnamefont {J.}~\bibnamefont {Schubert}},
  \bibinfo {author} {\bibfnamefont {J.~W.}\ \bibnamefont {Freeland}}, \bibinfo
  {author} {\bibfnamefont {D.~A.}\ \bibnamefont {Muller}}, \bibinfo {author}
  {\bibfnamefont {C.~J.}\ \bibnamefont {Fennie}}, \bibinfo {author}
  {\bibfnamefont {P.}~\bibnamefont {Schiffer}}, \bibinfo {author}
  {\bibfnamefont {V.}~\bibnamefont {Gopalan}}, \bibinfo {author} {\bibfnamefont
  {E.}~\bibnamefont {Johnston-Halperin}}, \ and\ \bibinfo {author}
  {\bibfnamefont {D.~G.}\ \bibnamefont {Schlom}},\ }\href
  {https://doi.org/10.1038/nature09331} {\bibfield  {journal} {\bibinfo
  {journal} {Nature}\ }\textbf {\bibinfo {volume} {466}},\ \bibinfo {pages}
  {954} (\bibinfo {year} {2010})}\BibitemShut {NoStop}%
\bibitem [{\citenamefont {Akamatsu}\ \emph {et~al.}(2011)\citenamefont
  {Akamatsu}, \citenamefont {Kumagai}, \citenamefont {Oba}, \citenamefont
  {Fujita}, \citenamefont {Murakami}, \citenamefont {Tanaka},\ and\
  \citenamefont {Tanaka}}]{PhysRevB.83.214421}%
  \BibitemOpen
  \bibfield  {author} {\bibinfo {author} {\bibfnamefont {H.}~\bibnamefont
  {Akamatsu}}, \bibinfo {author} {\bibfnamefont {Y.}~\bibnamefont {Kumagai}},
  \bibinfo {author} {\bibfnamefont {F.}~\bibnamefont {Oba}}, \bibinfo {author}
  {\bibfnamefont {K.}~\bibnamefont {Fujita}}, \bibinfo {author} {\bibfnamefont
  {H.}~\bibnamefont {Murakami}}, \bibinfo {author} {\bibfnamefont
  {K.}~\bibnamefont {Tanaka}}, \ and\ \bibinfo {author} {\bibfnamefont
  {I.}~\bibnamefont {Tanaka}},\ }\href {\doibase 10.1103/PhysRevB.83.214421}
  {\bibfield  {journal} {\bibinfo  {journal} {Phys. Rev. B}\ }\textbf {\bibinfo
  {volume} {83}},\ \bibinfo {pages} {214421} (\bibinfo {year}
  {2011})}\BibitemShut {NoStop}%
\bibitem [{\citenamefont {Ranjan}\ \emph {et~al.}(2007)\citenamefont {Ranjan},
  \citenamefont {Nabi},\ and\ \citenamefont {Pentcheva}}]{Ranjan_2007}%
  \BibitemOpen
  \bibfield  {author} {\bibinfo {author} {\bibfnamefont {R.}~\bibnamefont
  {Ranjan}}, \bibinfo {author} {\bibfnamefont {H.~S.}\ \bibnamefont {Nabi}}, \
  and\ \bibinfo {author} {\bibfnamefont {R.}~\bibnamefont {Pentcheva}},\ }\href
  {\doibase 10.1088/0953-8984/19/40/406217} {\bibfield  {journal} {\bibinfo
  {journal} {Journal of Physics: Condensed Matter}\ }\textbf {\bibinfo {volume}
  {19}},\ \bibinfo {pages} {406217} (\bibinfo {year} {2007})}\BibitemShut
  {NoStop}%
\bibitem [{\citenamefont {Birol}\ and\ \citenamefont
  {Fennie}(2013)}]{PhysRevB.88.094103}%
  \BibitemOpen
  \bibfield  {author} {\bibinfo {author} {\bibfnamefont {T.}~\bibnamefont
  {Birol}}\ and\ \bibinfo {author} {\bibfnamefont {C.~J.}\ \bibnamefont
  {Fennie}},\ }\href {\doibase 10.1103/PhysRevB.88.094103} {\bibfield
  {journal} {\bibinfo  {journal} {Phys. Rev. B}\ }\textbf {\bibinfo {volume}
  {88}},\ \bibinfo {pages} {094103} (\bibinfo {year} {2013})}
   \BibitemShut
  {NoStop}%
\bibitem [{\citenamefont {Li}\ \emph {et~al.}(2014)\citenamefont {Li},
  \citenamefont {Zhou}, \citenamefont {Yan}, \citenamefont {Mandrus},\ and\
  \citenamefont {Keppens}}]{1.4902137}%
  \BibitemOpen
  \bibfield  {author} {\bibinfo {author} {\bibfnamefont {L.}~\bibnamefont
  {Li}}, \bibinfo {author} {\bibfnamefont {H.}~\bibnamefont {Zhou}}, \bibinfo
  {author} {\bibfnamefont {J.}~\bibnamefont {Yan}}, \bibinfo {author}
  {\bibfnamefont {D.}~\bibnamefont {Mandrus}}, \ and\ \bibinfo {author}
  {\bibfnamefont {V.}~\bibnamefont {Keppens}},\ }\href {\doibase
  10.1063/1.4902137} {\bibfield  {journal} {\bibinfo  {journal} {APL
  Materials}\ }\textbf {\bibinfo {volume} {2}},\ \bibinfo {pages} {110701}
(\bibinfo {year} {2014})}\BibitemShut {NoStop}%
\bibitem [{\citenamefont {Roy}\ \emph {et~al.}(2018)\citenamefont {Roy},
  \citenamefont {Khan},\ and\ \citenamefont {Mandal}}]{PhysRevB.98.134428}%
  \BibitemOpen
  \bibfield  {author} {\bibinfo {author} {\bibfnamefont {S.}~\bibnamefont
  {Roy}}, \bibinfo {author} {\bibfnamefont {N.}~\bibnamefont {Khan}}, \ and\
  \bibinfo {author} {\bibfnamefont {P.}~\bibnamefont {Mandal}},\ }\href
  {\doibase 10.1103/PhysRevB.98.134428} {\bibfield  {journal} {\bibinfo
  {journal} {Phys. Rev. B}\ }\textbf {\bibinfo {volume} {98}},\ \bibinfo
  {pages} {134428} (\bibinfo {year} {2018})}\BibitemShut {NoStop}%
\bibitem [{\citenamefont {Anisimov}\ \emph {et~al.}(1991)\citenamefont
  {Anisimov}, \citenamefont {Zaanen},\ and\ \citenamefont
  {Andersen}}]{PhysRevB.44.943}%
  \BibitemOpen
  \bibfield  {author} {\bibinfo {author} {\bibfnamefont {V.~I.}\ \bibnamefont
  {Anisimov}}, \bibinfo {author} {\bibfnamefont {J.}~\bibnamefont {Zaanen}}, \
  and\ \bibinfo {author} {\bibfnamefont {O.~K.}\ \bibnamefont {Andersen}},\
  }\href {\doibase 10.1103/PhysRevB.44.943} {\bibfield  {journal} {\bibinfo
  {journal} {Phys. Rev. B}\ }\textbf {\bibinfo {volume} {44}},\ \bibinfo
  {pages} {943} (\bibinfo {year} {1991})}\BibitemShut {NoStop}
  \end{thebibliography}
\end{document}